\documentclass[sigconf,screen,nonacm]{acmart}
\AtBeginDocument{%
  }

\setcopyright{cc}
\setcctype[4.0]{by}
\copyrightyear{2026}

\usepackage{multirow}
\usepackage{enumitem}

\newcommand{\OCR}{{\textit{OpenCodeReview}}}

\newcommand{\OCRLogo}{\includegraphics[height=1.8em]{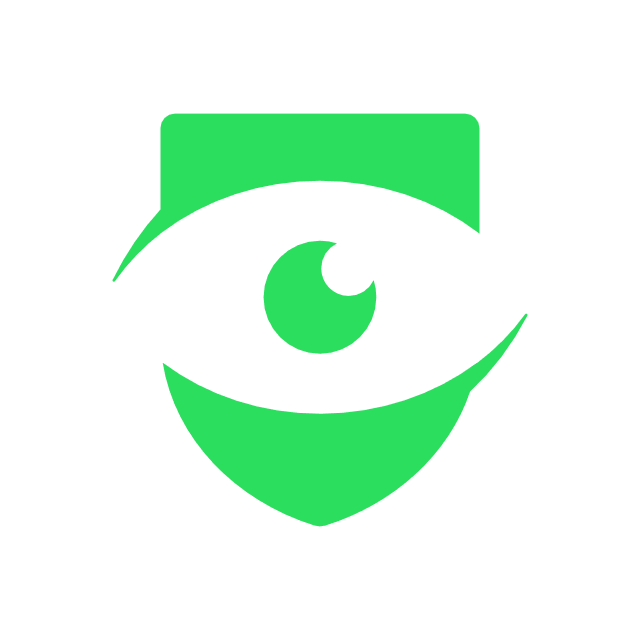}}

\begin{document}

\title{\raisebox{-0.35\height}{\OCRLogo}\hspace{0.1em}\textit{OpenCodeReview}: Determinism over Non-Determinism for Cost-Effective Agent-Based Code Review}

\author{Zhengfeng Li}
\affiliation{%
  \institution{Alibaba Group}
  \city{Hangzhou}
  \state{Zhejiang}
  \country{China}}
\email{lizhengfeng.lzf@alibaba-inc.com}

\author{Lei Zhang}
\affiliation{%
  \institution{Nanjing University}
  \city{Nanjing}
  \state{Jiangsu}
  \country{China}}
\email{602025320025@smail.nju.edu.cn}

\author{Xianwei Wu}
\affiliation{%
  \institution{Nanjing University}
  \city{Nanjing}
  \state{Jiangsu}
  \country{China}}
\email{652024320006@smail.nju.edu.cn}

\author{Zhengqi Zhuang}
\affiliation{%
  \institution{Nanjing University}
  \city{Nanjing}
  \state{Jiangsu}
  \country{China}}
\email{zephyrqz@163.com}

\author{Yingjie Xu}
\affiliation{%
  \institution{Peking University}
  \city{Beijing}
  \country{China}}
\email{xuyingjie@stu.pku.edu.cn}

\author{Boge Wang}
\affiliation{%
  \institution{Alibaba Group}
  \city{Hangzhou}
  \state{Zhejiang}
  \country{China}}
\email{bogw.wbg@alibaba-inc.com}

\author{Shaofei Zhu}
\affiliation{%
  \institution{Alibaba Group}
  \city{Hangzhou}
  \state{Zhejiang}
  \country{China}}
\email{zhushaofei.zsf@alibaba-inc.com}

\author{Chuan Wang}
\affiliation{%
  \institution{Alibaba Group}
  \city{Hangzhou}
  \state{Zhejiang}
  \country{China}}
\email{nashui.wc@alibaba-inc.com}

\author{Peng Zhao}
\affiliation{%
  \institution{Alibaba Group}
  \city{Hangzhou}
  \state{Zhejiang}
  \country{China}}
\email{zhuyun.zp@alibaba-inc.com}

\author{Xinyu Zheng}
\affiliation{%
  \institution{Alibaba Group}
  \city{Hangzhou}
  \state{Zhejiang}
  \country{China}}
\email{yuxin.zxy@alibaba-inc.com}

\author{Guoping Rong}
\affiliation{%
  \institution{Nanjing University}
  \city{Nanjing}
  \state{Jiangsu}
  \country{China}}
\email{ronggp@nju.edu.cn}

\renewcommand{\shortauthors}{Li et al.}

\begin{abstract}
LLM-based code review agents promise scalable, always-on review, yet current systems suffer from two intertwined weaknesses: (1) \textit{non-determinism}—unbounded tool use and sprawling action spaces make review outcomes unstable across runs, and (2) \textit{context locality}—the reviewer's effective access remains bounded to the diff itself, capping the depth of discoverable issues. Both weaknesses give rise to three challenges: misaligned context retrieval that either pollutes or under-supplies the agent's context, a coherence-efficiency trade-off in scaling to multi-file pull requests, and hallucinated comments that erode trust and impose a vetting cost. To address these challenges, we introduce \OCR, an LLM-based code review agent built on the principle of deterministic engineering for uncertain agents: rather than granting the agent maximal freedom, we inject determinism at three deliberate points in the review pipeline. \textbf{Rule-Guided Dispatch} uses a multi-layer rule system to deterministically select files and review criteria, eliminating the variability of agent-driven file triage. \textbf{Grounded File Review} replaces free-form exploration with a curated, review-specific tool set with bounded outputs exposed through a ReAct loop, while file-level parallel SubAgents balance context coherence against execution efficiency and recover cross-file dependencies on demand. \textbf{Independent Reflection} introduces a falsification-first filter under an asymmetric information boundary—the reflector sees only the diff, not the agent's tool-augmented exploration—that removes hallucinated comments without the self-reinforcing bias of intrinsic self-critique, improving precision while preserving recall. On AACR-Bench (i.e., 200 real-world PRs, 10 languages, 1{,}505 expert-verified comments), \OCR\ consistently outperforms mainstream coding agents (e.g., Claude Code and Codex) across six LLM backends, achieving up to 2.17$\times$ higher SEM-F1 (25.10\% vs.\ 11.57\% for the same model under Claude Code) while consuming 5--15$\times$ fewer tokens---demonstrating that depth, reliability, and efficiency can be achieved simultaneously.
We open-source \OCR\ at \url{https://github.com/alibaba/open-code-review} to foster further research and adoption of deterministic engineering for agent-based code review.
\end{abstract}


\begin{teaserfigure}
 \includegraphics[width=\textwidth]{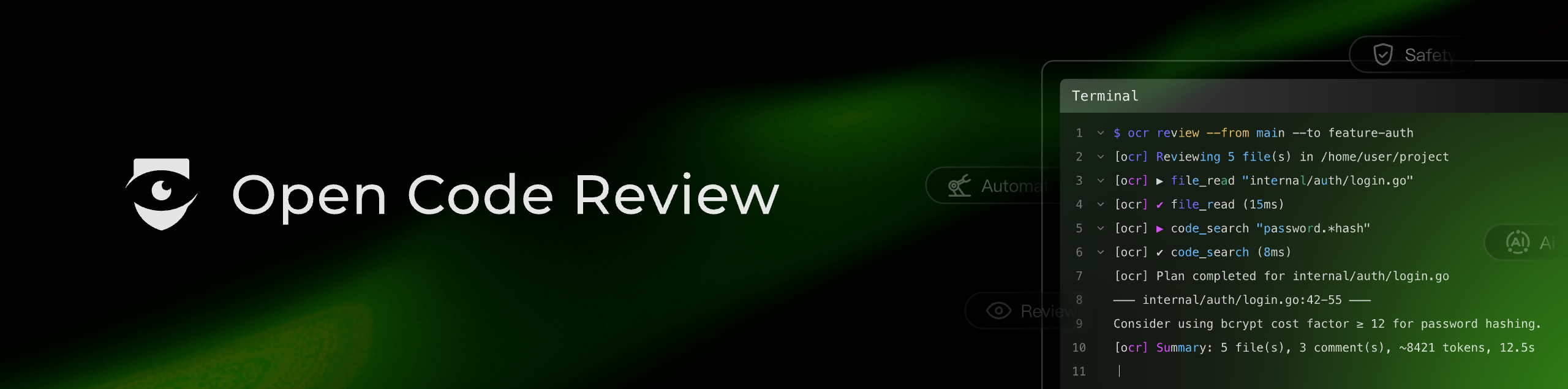}
  \caption{Banner of OpenCodeReview.}
  \Description{Banner of OpenCodeReview}
  \label{fig:teaser}
\end{teaserfigure}


\maketitle

\section{Introduction}

Code review is a cornerstone of modern software quality assurance, through which developers collaboratively inspect changes to detect defects, improve maintainability, and propagate knowledge~\cite{bacchelli2013modern,rigby2013peer}. Yet the high cost of manual review—reviewer time, cognitive load, and turnaround delay—has long constrained its adoption at scale: developers spend a substantial fraction of their time on review, and pull-request volume in large projects frequently outpaces human capacity~\cite{sadowski2018modern}. This tension has motivated sustained efforts to automate the review process.

Automated code review aims to augment or replace human reviewers with methods that improve review efficiency and scalability~\cite{li2022codereviewer}. Early approaches, from static checkers to learning-based comment generators, offered limited coverage and fluency. The rapid progress of large language models (LLMs) has reshaped this landscape by enabling fluent review feedback at near-zero marginal cost, transforming scalable, always-on reviewers from a long-standing aspiration into a realistic prospect. Consequently, LLM-based code review has attracted broad attention, with a fast-growing body of work at top software engineering and AI venues.

Existing explorations fall into three lines: strengthening the model's reviewing ability through pre-training or fine-tuning~\cite{li2022codereviewer,lu2023llamareviewer,thongtan2024finetuning,haider2024prompting}, augmenting the input with retrieved context~\cite{zhang2025laura,icoz2026contextaware}, and decomposing review across multiple agents~\cite{tang2024codeagent,sharanarthi2025multiagent} (Section~\ref{sec:related_work_code_review}). In parallel, industry has also embraced this trend: mainstream coding agents such as Claude Code\footnote{\url{https://www.anthropic.com/claude-code}}, Codex\footnote{\url{https://openai.com/index/codex/}}, Cursor\footnote{\url{https://www.cursor.com}}, and OpenCode\footnote{\url{https://github.com/sst/opencode}} now ship built-in code review, signaling that LLM-based review has moved from prototypes to deployed products.

Despite this progress, feedback that sounds plausible is not yet feedback that is genuinely useful. A meaningful comment must reason about the change in the context of the entire repository (i.e., how it interacts with callers, callees, shared data structures, and project conventions) rather than the changed lines alone. Yet across all three lines, the reviewer's effective context is ultimately bounded to the diff and its surroundings (Section~\ref{sec:related_work_code_review}), capping the depth of discoverable issues: cross-file control- or data-flow bugs, regressions that break distant callers, and repository-wide convention violations remain largely out of reach. The emerging paradigm of LLM-based agent systems that interleave reasoning, tool use, and multi-step exploration\cite{yao2023react} (Section~\ref{sec:related_work_llm_agent})—offers a path to break this bound by actively gathering repository-level context on demand, as shown in autonomous program repair and issue resolution\cite{zhang2024autocoderover,yang2024sweagent}. However, adapting this paradigm to code review exposes three coupled challenges.

\textbf{First, the scope and strategy of context retrieval remain a fundamental tension}. An agent that arbitrarily expands its context risks diluting the signal with irrelevant code (i.e., context pollution), while one that retrieves too conservatively collapses back to the locality-bounded regime. Existing agents typically use a fixed, hand-designed set of retrieval tools borrowed from program repair or code search, whose patterns are seldom validated against how human reviewers navigate a repository~\cite{zhang2023repocoder,zhang2024autocoderover}, so the tools and the genuine information needs of review are often misaligned. A tempting remedy is a general-purpose \emph{bash} tool for free repository exploration; while this maximizes context-gathering freedom, unbounded tool use inflates the context and token budget with redundant output and the vast action space introduces pronounced non-determinism, making review outcomes far less stable and reproducible across runs~\cite{wang2025efficient,fan2025sweeffi,yagubyan2026consistency}.

\textbf{Second, scaling agent-based review to a multi-file pull request forces a trade-off between context coherence and execution efficiency}. A single monolithic agent reviewing the entire changeset under one shared context preserves a holistic view, but its cost grows prohibitively with change size and a sprawling context window dilutes the signal. At the other extreme, overly fine partitioning (e.g., per hunk or per function review) maximizes parallelism but fragments a coherent change across isolated agents, inflating coordination overhead and obscuring the change as a whole. Prior multi-agent systems adopt either centralized, conversational designs whose coordination limits scalability~\cite{tang2024codeagent}, or role decompositions inherited from development workflows that do not map onto review~\cite{qian2024chatdev,hong2024metagpt}. How to choose a granularity that preserves change coherence while staying tractably parallel and lets each reviewer still recover cross-file dependencies remains open.

\textbf{Third, the autonomy that makes agents powerful but also unreliable}. Multi-step agents tend to hallucinate by asserting non-existent facts or inventing unsupported issues, and this instability compounds over long agent loops~\cite{shinn2023reflexion,ji2023survey}. In code review scenario, hallucinated comments are especially damaging: a reviewer that interleaves genuine and fabricated findings erodes trust and imposes a vetting cost that can negate the savings automation was meant to provide. Existing mitigations rely on either intrinsic self-reflection, where the same model critiques its own output and inherits the same biases~\cite{shinn2023reflexion,madaan2023selfrefine}, or heavyweight program-analysis validators that verify a narrow class of findings but cannot adjudicate broader semantic review comments~\cite{guo2025repoaudit} (Section~\ref{sec:related_work_reflection}). A mechanism that intercepts hallucinated comments across different types, without sacrificing issue breadth, remains missing.

To address these challenges, we present \OCR, an LLM-based code review agent built on the principle of \emph{deterministic engineering for uncertain agents}: rather than granting the agent maximal freedom, we inject determinism at three points in the review pipeline to ensure both the depth of issues surfaced and the reliability of each comment. \OCR\ is built around three core designs:

\begin{enumerate}[leftmargin=*]
  \item \textbf{Rule-Guided Dispatch} uses a multi-layer rule system to deterministically select files and review criteria, eliminating the variability of agent-driven file triage. Rather than letting the agent decide which files merit attention, rule-driven dispatch ensures that the same PR always yields the same file and criterion assignment, providing a reproducible starting point for subsequent review.

  \item \textbf{Grounded File Review} replaces free-form exploration with a curated, review-specific tool set with bounded outputs exposed through a ReAct loop, while file-level parallel SubAgents balance context coherence against execution efficiency and recover cross-file dependencies on demand. Grounding tool design in a data-driven analysis of the context-retrieval patterns that expert reviewers use on real diffs aligns the agent's information-gathering with the genuine needs of code review, uncovering deep, repository-spanning issues while keeping retrieved context informative and cost-bounded.

  \item \textbf{Independent Reflection} introduces a falsification-first filter under an asymmetric information boundary: the reflector sees only the diff, not the agent's tool-augmented exploration, removing hallucinated comments without the self-reinforcing bias of intrinsic self-critique~\cite{shinn2023reflexion,madaan2023selfrefine}. By design, the reflector only filters rather than generates, improving precision while preserving recall.
\end{enumerate}

We evaluate \OCR\ on AACR-Bench~\cite{zhang2026aacrbench}, a multilingual benchmark comprising 200 real-world PRs and 1{,}505 expert-verified comments, comparing it with Claude Code and Codex under six LLM backends. \OCR\ achieves the highest SEM-F1 across all configurations, with the top result (25.10\% under Claude-4.6-Opus) outperforming the same model under Claude Code (11.57\%) by 2.17$\times$, while reducing token consumption by 5--15$\times$. These results confirm that structured, deterministic engineering yields both deeper review and higher reliability at substantially lower cost.

In summary, we make the following contributions:

\begin{enumerate}
  \item We propose \OCR, an LLM-based code review agent built on the principle of \emph{deterministic engineering for uncertain agents}, which injects determinism at three points in the review pipeline to ensure both depth and reliability of review outcomes.
  \item We conduct a comprehensive evaluation on AACR-Bench across six LLM backends, demonstrating that \OCR\ consistently outperforms mainstream coding agents with up to 2.17$\times$ higher SEM-F1 and 5--15$\times$ fewer tokens.
  \item We open-source \OCR\ to facilitate reproducibility and foster future research on deterministic agent design for code review.
\end{enumerate}

\section{Background and Related Work}

\label{sec:related_work}

This section surveys the landscape that motivates \OCR. We first review the evolution of the LLM-based code review, then cover the LLM agents for software engineering, and the reflection mechanisms for improving agent reliability.

\subsection{LLM-Based Code Review}

\label{sec:related_work_code_review}

Modern code review is a widely adopted practice in which developers inspect pull requests (PRs) to detect defects, improve maintainability, and share knowledge~\cite{bacchelli2013modern,rigby2013peer}. At Google, code review has been identified as a primary bottleneck on developer throughput~\cite{sadowski2018modern}, motivating sustained efforts toward automation. The rapid progress of large language models (LLMs) has reshaped this landscape by enabling fluent review feedback at near-zero marginal cost, transforming scalable, always-on reviewers from a long-standing aspiration into a realistic prospect. We organize LLM-based code review methods along three lines: review-oriented model training, retrieval augmentation, and agentic code review.

\paragraph{Review-oriented model training.}
This line adapts LLMs to code review through task-specific fine-tuning. LLaMA-Reviewer~\cite{lu2023llamareviewer} and follow-ups~\cite{thongtan2024finetuning,haider2024prompting} fine-tune open-source LLMs on review comment data to improve comment generation quality. \citet{yu2025finetuning} further show that fine-tuning can simultaneously enhance the accuracy and comprehensibility of generated comments. More recently, MelcotCR~\cite{yu2025melcotcr} proposes a maximum entropy regulated long chain-of-thought fine-tuning approach that trains LLMs to analyze multiple dimensions of code review simultaneously, enabling a 14B model to match the performance of a 671B model. Despite these advances, the reviewing knowledge is ultimately encoded in model weights: at inference time, the model cannot consult the target repository and remains bounded to the diff, limiting its ability to surface issues that require cross-file reasoning.

\paragraph{Retrieval augmentation.}
This line augments the model input with retrieved context to partially overcome the locality of diff-only review. AUGER~\cite{du2022auger} retrieves similar code snippets to seed comment generation. RAG-Reviewer~\cite{hong2025ragreviewer} unifies generation- and IR-based methods by conditioning on top-$k$ similar code--review pairs, improving comment quality especially for low-frequency tokens. LAURA~\cite{zhang2025laura} enriches input with PR metadata, AST-expanded context, and historically similar reviews retrieved via CodeT5+ embeddings. \citet{icoz2026contextaware} ground comments in retrieved snippets through a context-aware RAG pipeline. Sun et al.~\cite{sun2026issuelist} systematically compare neighboring, LSP-based semantic, and IR-based similar co-change context, finding that combining issue-list review with context augmentation substantially improves review coverage. While these methods broaden the effective context, they do so only along axes the retriever is designed for: a fixed retriever cannot adaptively trace the evidence chains that human reviewers follow (e.g., ``\textit{who calls this changed function, and what do they expect?}''), nor can it adapt its strategy to the information needs of a given change.

\paragraph{Agentic code review.}
This line adopts an agent-based paradigm for code review. Research systems decompose review into specialized roles or supervisor--worker designs with verification agents~\cite{tang2024codeagent,sharanarthi2025multiagent}. Industry has meanwhile shipped review functionality in mainstream coding agents, including Cursor's BugBot, Claude Code, Codex, and OpenCode, each offering diff-based or sandboxed-agent review, but without purpose-built context tools or multi-file parallelism. A recent empirical study of $3{,}109$ PRs reports that code-review-agent-only reviews achieve a $45.2\%$ merge rate (vs.\ $68.4\%$ for human-only), with $60\%$ of rejected agent-only PRs exhibiting a signal-to-noise ratio below $30\%$~\cite{chowdhury2026empirical}, evidence that current deployed agents produce noisy, low-quality feedback. These systems share three limitations: (1)~centralized coordination constrains scalability, as a single conversation thread cannot efficiently process large, multi-file changes; (2)~role decompositions borrowed from development workflows (e.g., coder, tester) do not align with the structure of review; and (3)~none employ purpose-built context tools, leaving the agent to either rely on the diff alone or explore the repository with general-purpose, unbounded tools. 

To address these limitations, we propose \OCR, an LLM-based code review agent built on the principle of deterministic engineering for uncertain agents, with three core designs: \textbf{Rule-Guided Dispatch} (deterministic file and criterion selection), \textbf{Grounded File Review} (curated context tools and file-level parallel SubAgents), and \textbf{Independent Reflection} (falsification-based comment filtering).

\subsection{LLM Agents for Software Engineering}

\label{sec:related_work_llm_agent}

An LLM agent places the model in a loop in which it reasons about state, invokes tools to obtain observations, and selects the next action; the ReAct paradigm~\cite{yao2023react} formalizes this as a Thought-Action-Observation trajectory. Software engineering is a natural application domain, where a repository is a queryable environment, and an agent can search definitions, inspect callees, and run analyses to gather evidence. This insight has driven rapid progress on bug repair and issue resolution domains. SWE-agent~\cite{yang2024sweagent} resolves GitHub issues via a custom agent--computer interface, demonstrating that tool-interface design materially affects performance. AutoCodeRover~\cite{zhang2024autocoderover} decomposes repair into context retrieval and patching over a spectrum of retrieval tools. RepoAudit~\cite{guo2025repoaudit} audits repositories and validates findings with program analysis. ChatDev~\cite{qian2024chatdev} and MetaGPT~\cite{hong2024metagpt} organize development into role-specialized multi-agent systems.

These works confirm that agents can navigate repositories effectively, informing our design of curated context tools for Grounded File Review. However, existing agent systems differ from code review in two key respects that motivate our file-level SubAgent partitioning: (1)~their objective is to produce a fix or verdict, whereas review must discover and articulate problems, requiring different tools and success criteria; and (2)~their organization relies on either a single agent or a development-role decomposition, neither of which naturally suits multi-file PR review. Our file-level partitioning is purpose-built for review, directly addressing the coherence-efficiency trade-off between context coherence and execution efficiency.

\subsection{Agent Reliability and Reflection Mechanisms}

\label{sec:related_work_reflection}

The unreliability of LLM agent outputs stems from two sources: the hallucination inherent to the model and the non-determinism introduced by the agent's tool-use trajectory. Both make autonomous review unsafe to deploy without safeguards, and each demands a different mitigation strategy.

Regarding hallucination, LLM agents tend to assert unsupported facts, and this tendency compounds over long tool-calling loops~\cite{ji2023survey}. Existing mitigations fall into two camps: (1)~\textit{Intrinsic self-reflection} (e.g., Reflexion~\cite{shinn2023reflexion}, Self-Refine~\cite{madaan2023selfrefine}) has the same model critique its own output, inheriting the biases that caused the error; and (2)~\textit{External validation} (e.g., RepoAudit's program-analysis checks~\cite{guo2025repoaudit}) is reliable only for the narrow facts the analyzer verifies and cannot adjudicate broader semantic comments. Our Independent Reflection module addresses hallucination by decoupling the reflector from the reviewing agent, avoiding the self-reinforcing bias of intrinsic reflection. By only filtering and never generating, it improves precision without constraining issue breadth.

Regarding non-determinism, free-form tools such as \texttt{bash} inflate the context budget, producing the ``\textit{token snowball}'' effect~\cite{wang2025efficient,fan2025sweeffi}, and vast action spaces also introduce variability across runs~\cite{yagubyan2026consistency}, motivating the curated, review-specific tool set in our Grounded File Review module rather than free-form exploration.

\begin{figure*}[ht]
    \centering
    \includegraphics[width=1.0\linewidth]{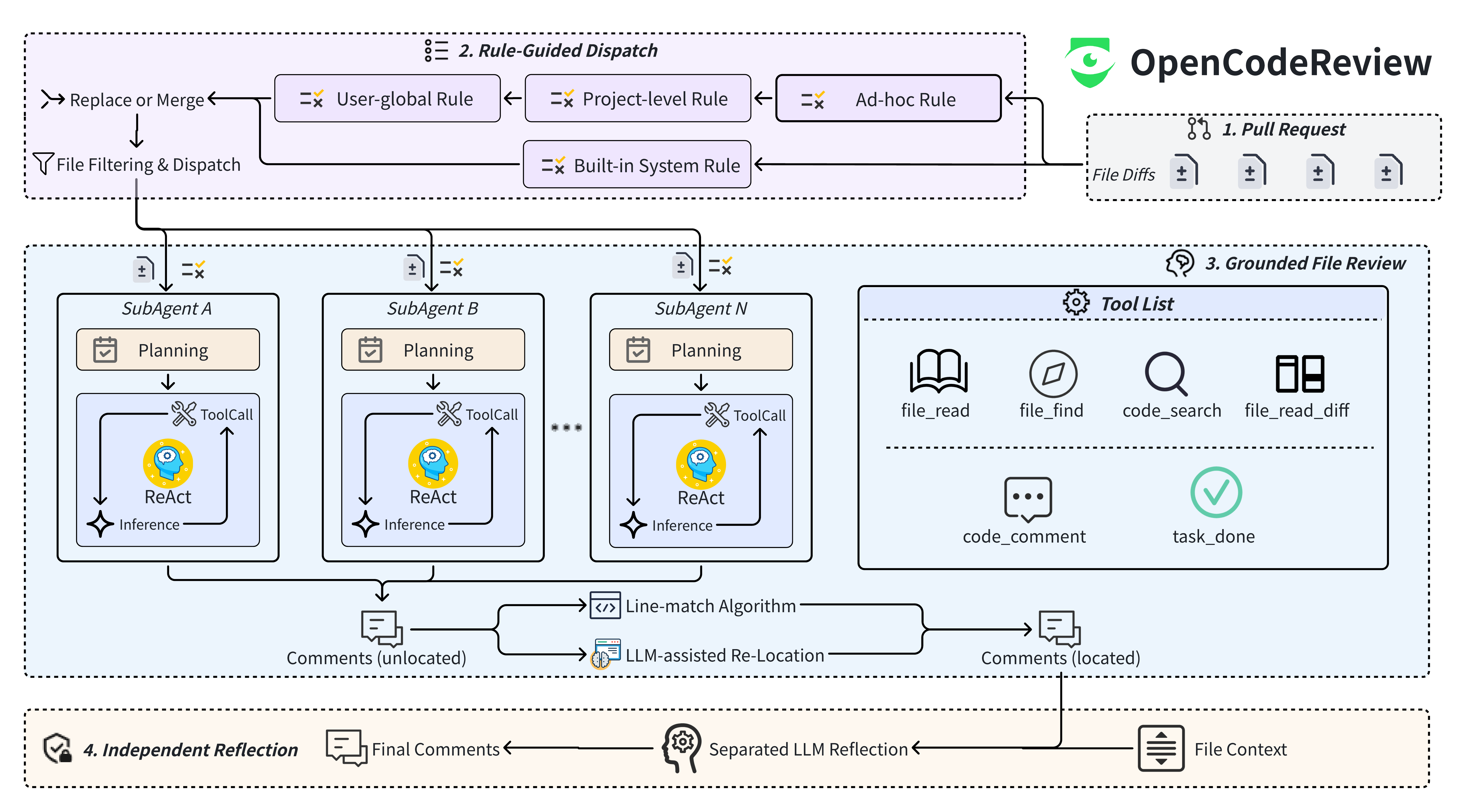}
    \caption{Overview of \OCR. The pipeline proceeds in three stages. (1)~\textbf{Rule-Guided Dispatch}: given a PR, the system resolves applicable rules from a four-tier chain (built-in, user-global, project-level, and ad-hoc) for each changed file, filters files by extension, user include/exclude patterns, and size, and dispatches one SubAgent per file in parallel. (2)~\textbf{Grounded File Review}: each SubAgent executes a ReAct loop with the file's diff, the resolved rule, and a curated tool set (\texttt{file\_read}, \texttt{file\_find}, \texttt{code\_search}, \texttt{file\_read\_diff}, \texttt{code\_comment}, \texttt{task\_done}), whose bounded outputs prevent context bloat while enabling cross-file exploration. Generated comments are resolved to precise line numbers via a multi-stage fallback mechanism. (3)~\textbf{Independent Reflection}: an external reflector examines each comment against the diff alone under an asymmetric information boundary, filtering those directly contradicted by diff evidence without the self-reinforcing bias of same-model critique.}
    \label{fig:ocr-overview}
\end{figure*}

\section{The \OCR\ Agent}

\subsection{Overview}

As established in Section~\ref{sec:related_work}, existing approaches to automated code review suffer from two intertwined weaknesses: (1) \textit{non-determinism}, where unbounded tool use and sprawling action spaces make agent behavior unstable across runs~\cite{wang2025efficient,fan2025sweeffi,yagubyan2026consistency}, and (2) \textit{context locality}, where the reviewer's effective access is ultimately bounded to the diff and its surroundings, capping the depth of discoverable issues. Both two weaknesses are related in practice: the free-form tools prevalent in agent-based systems (e.g., general-purpose shell access) are a key contributor to non-determinism, while the fixed retrievers dominant in retrieval-augmented methods are a key contributor to context locality. The root cause is a single design gap: no existing system constrains the agent's behavior along axes that are validated against the genuine information needs of code review.

\OCR\ is built on a design philosophy we call \emph{deterministic engineering for uncertain agents}. Rather than granting the agent maximal freedom and hoping it converges, we inject determinism at three points in the review pipeline, each addressing a specific source of non-determinism or locality identified in Section~\ref{sec:related_work}:

\begin{enumerate}[leftmargin=*]
  \item \textbf{Rule-Guided Dispatch} replaces ad-hoc file selection with a multi-layer rule system that deterministically decides which files to review and what review criteria to apply, eliminating the non-determinism of agent-driven file triage.
  \item \textbf{Grounded File Review} replaces free-form exploration with a curated, review-specific tool set exposed through a ReAct loop, constraining the agent's action space to operations that are both productive for review and bounded in their context footprint, addressing the ``\textit{token snowball}'' effect~\cite{fan2025sweeffi} while breaking the locality bound.
  \item \textbf{Independent Reflection} replaces intrinsic self-critique with an external falsification check that operates under an asymmetric information boundary, filtering hallucinated comments without the self-reinforcing bias of same-model reflection~\cite{shinn2023reflexion,madaan2023selfrefine}.
\end{enumerate}

Together, these three modules form a pipeline that is deterministic in its dispatch, grounded in its exploration, and reliable in its output. Figure~\ref{fig:ocr-overview} illustrates the overall architecture and the remainder of this section details each module.

\subsection{Rule-Guided Dispatch}

The first source of non-determinism in agent-based review is file selection: when asked to ``review the PR,'' an agent must decide which files merit attention and what criteria to apply. If left to the agent, this decision varies across runs and may overlook files governed by project-specific review standards. \OCR\ eliminates this variability by making file selection and criterion assignment rule-driven rather than agent-driven.

\subsubsection{The Rule Concept}

A rule in \OCR\ is a natural-language document specifying review criteria for a class of files. Rules serve a dual purpose: they determine which files are in scope (via glob-based path patterns) and what the SubAgent should check (via structured review checklists covering correctness, security, performance, maintainability, and test coverage). Each rule document is domain-specific. For example, a rule for a statically typed language may include checks for thread safety and common framework pitfalls, while a rule for CI configuration files focuses on workflow syntax and secret management. This ensures that the review criteria are tailored to the file type rather than generic.

\subsubsection{Multi-Layer Rule Resolution}

Rules are organized into a four-tier priority chain (Table~\ref{tab:rule-tiers}), each tier corresponding to a different scope of authority.

\begin{table*}[t]
  \centering
  \caption{Four-tier rule priority chain.}
  \label{tab:rule-tiers}
  \begin{tabular}{cll}
    \toprule
    \textbf{Tier} & \textbf{Source} & \textbf{Scope} \\
    \midrule
    1 (highest) & Ad-hoc rule provided at invocation time & Per-invocation \\
    2 & Project-level rule stored in the repository & Project-specific, version-controlled \\
    3 & User-global rule stored in the user's home directory & Cross-project \\
    4 (lowest) & Built-in system rules & Defaults per language/file type \\
    \bottomrule
  \end{tabular}
\end{table*}

At the base tier, the system ships a set of built-in rules covering major programming languages and configuration file formats. Path matching uses recursive glob patterns with brace expansion and case-insensitive comparison, evaluated in declaration order with first-match-wins semantics. This ensures that each file is deterministically mapped to the most specific applicable rule.

When a user defines a project or global rule, it can interact with the system rule in one of two modes:

\begin{itemize}[leftmargin=*]
  \item \textbf{Replace mode (default)}: the user rule entirely supersedes the system rule for matching paths.
  \item \textbf{Merge mode}: the system and user rule are concatenated into a single document, preserving language-specific defaults while appending project-specific criteria.
\end{itemize}

This layered design ensures that review criteria are deterministic: the same PR always yields the same rule assignment, while remaining adaptable in that projects can override or extend defaults without modifying the system.

\subsubsection{File Filtering and Dispatch}

Before dispatching SubAgents, \OCR\ applies a multi-stage filter to the changed files: (1) binary files are excluded; (2) user-defined exclude patterns are applied if configured; (3) if include patterns are defined, only matching files pass; (4) a built-in extension allowlist filters unsupported file types; and (5) default exclusion patterns remove test fixtures and generated code. Additionally, (6) files whose diff content exceeds $80\%$ of the model's context window are filtered to avoid context saturation.
For each remaining file, the system resolves the applicable rule text via the four-tier chain and constructs the SubAgent's input by injecting the file's diff, the resolved rule, the list of other changed files, and any user-provided background information into a prompt template. Each SubAgent is then launched as an independent concurrent task, enabling file-level parallelism.

This design ensures that the \textit{what} (which files) and the \textit{how} (what criteria) of review are fully determined by rules rather than agent discretion, eliminating a major source of run-to-run variability while preserving the flexibility to adapt to project conventions.

\subsection{Grounded File Review}

Once a SubAgent is dispatched with a file's diff and applicable rules, it must explore the repository to gather the context needed to identify deep, cross-file issues without the non-determinism and context bloat of free-form tools. \OCR\ achieves this through a ReAct-based agent loop with a curated tool set, designed to ground the agent's reasoning in evidence while bounding its context footprint.

\subsubsection{The agent Loop}

Each SubAgent follows a standard ReAct loop~\cite{yao2023react}: in each iteration, the LLM produces a thought (reasoning about the current state) and selects an action (a tool call); the tool's output is returned as an observation and appended to the conversation history. The loop continues until the agent invokes \texttt{task\_done} or reaches a maximum iteration bound. We set the default bound to 30 as a practical trade-off: it is high enough to accommodate typical cross-file exploration (e.g., tracing callers, inspecting callees, and checking related tests), yet low enough to serve as a hard ceiling against runaway loops. Two additional control mechanisms further prevent the loop from degenerating before this bound is reached:

\begin{itemize}[leftmargin=*]
  \item \textbf{Empty-round detection}. If the agent produces no tool calls for three consecutive rounds, it is prompted to either act or terminate, preventing infinite stalling.
  \item \textbf{Context compression}. As the conversation grows, \OCR\ monitors token usage against the model's context window. At $60\%$ utilization, an asynchronous background compression is triggered; at $80\%$, a synchronous compression forcibly summarizes the middle region of the conversation history into a structured summary while preserving the system prompt and the most recent interactions. This three-region strategy (frozen--compress--active) mitigates the token snowball effect~\cite{fan2025sweeffi,wang2025efficient}.
\end{itemize}

\subsubsection{Curated Context Tools}

Rather than exposing a general \texttt{bash} shell, which maximizes flexibility at the cost of inflated action space and context budget~\cite{yagubyan2026consistency}, \OCR\ provides a fixed set of six review-specific tools, each with a bounded output. Table~\ref{tab:tools} summarizes the tool set.

\begin{table*}[t]
  \centering
  \caption{Curated tool set for grounded file review.}
  \label{tab:tools}
  \begin{tabular}{lll}
    \toprule
    \textbf{Tool} & \textbf{Purpose} & \textbf{Bounded Output} \\
    \midrule
    \texttt{file\_read} & Read a file's content by path and optional line range & Max 500 lines per call \\
    \texttt{file\_find} & Locate files by name keyword & Max 100 results \\
    \texttt{code\_search} & Search for text patterns across the repository & Max 100 matches, 10s timeout \\
    \texttt{file\_read\_diff} & View the diff of another changed file & Returns the pre-computed diff \\
    \texttt{code\_comment} & Submit a review comment & Parsed and collected asynchronously \\
    \texttt{task\_done} & Signal task completion & Terminates the loop \\
    \bottomrule
  \end{tabular}
\end{table*}

This tool set is motivated by the observation that tool-interface design materially affects agent performance~\cite{yang2024sweagent}. Each tool corresponds to a distinct information need that a human reviewer would fulfill: reading the full context of a changed function (\texttt{file\_read}), locating a definition or test (\texttt{file\_find}), tracing callers of a changed API (\texttt{code\_search}), understanding how a change interacts with concurrent modifications (\texttt{file\_read\_diff}), and recording a finding (\texttt{code\_comment}). Among these, \texttt{file\_read\_diff} is unique to the multi-file PR setting, and a strict focus rule in the system prompt instructs the SubAgent that findings from other files must not become the subject of comments, as only the current file's diff is under review.

The bounded output of each tool is a deliberate constraint: by capping results and enforcing timeouts, the system prevents any single tool call from saturating the context window, keeping the agent's behavior predictable and cost-bounded without sacrificing the ability to trace cross-file dependencies.

\subsubsection{Comment Generation and Line Resolution}

When the agent identifies an issue, it invokes \texttt{code\_comment} with the comment text, a code snippet for localization (\texttt{existing\_code}), and an optional suggested fix (\texttt{suggestion\_code}). Since the agent may produce imprecise or incomplete code snippets, \OCR\ resolves line numbers through a three-stage fallback rather than trusting the agent's output directly. In the first stage, the system attempts to match \texttt{existing\_code} against the new-side hunks of the diff, which yields precise line anchors within the changed region. If the match fails, the second stage searches the full file content, handling cases where the snippet references unchanged lines outside the diff. If both stages fail, the third stage invokes an LLM-assisted relocation step that re-extracts the referencing snippet from the surrounding context. This multi-stage fallback ensures that comments are accurately anchored even when the agent's localization is unreliable.

\subsection{Independent Reflection}

The autonomy that makes the ReAct loop effective but also makes it prone to hallucination: the agent may assert issues that the diff evidence does not support or directly contradicts~\cite{ji2023survey}. Existing mitigations, as discussed in Section~\ref{sec:related_work_reflection}, either rely on intrinsic self-reflection (inheriting the same biases) or external program-analysis validators (covering only narrow fact types). \OCR\ introduces a reflection mechanism that is independent in its information boundary yet broad in its coverage.

\subsubsection{Asymmetric Information Boundary}

The reflection module is invoked after a SubAgent completes its ReAct loop. It receives two inputs: the file's diff and the list of comments produced by the SubAgent. Critically, the reflector operates under an asymmetric information boundary: unlike the SubAgent, which had access to the full repository via tools, the reflector sees only the diff. This design is deliberate. The SubAgent may have formed its comments using context gathered from other files, context that the reflector cannot verify. Rather than attempting to re-derive this context, the reflector's task is narrowly scoped: filter only those comments that are directly contradicted by evidence within the diff itself.

This asymmetry is the key to the module's independence. Intrinsic self-reflection fails because the same model, reviewing its own output with the same context, tends to confirm its prior conclusions~\cite{shinn2023reflexion,madaan2023selfrefine}. External validators fail because they cover only the narrow facts that program analyzers can check. The reflection module occupies a middle ground: it uses the same LLM, but its information boundary is different. It sees less than the agent, not more, which breaks the self-reinforcing bias without restricting coverage to analytically verifiable fact types.

\subsubsection{Falsification, Not Verification}

The reflector's guiding principle is falsification rather than verification, inspired by the epistemological principle that claims should be tested by attempting to disprove them rather than confirm them~\cite{huang2025popper}. Concretely, the reflector performs a two-step evaluation for each comment:

\begin{itemize}[leftmargin=*]
  \item \textbf{Fact check (veto rule)}: The reflector examines whether the diff contains direct counter-evidence to the comment's key claim. If the diff contradicts the comment, the comment is flagged for removal. If the comment references context not visible in the diff (e.g., logic in other files, runtime behavior), the reflector does not flag it, as the SubAgent may have valid evidence from its tool-augmented exploration.
  \item \textbf{Issue classification}: For comments whose diff-visible facts are accurate, the reflector checks whether the comment mischaracterizes the code, e.g., flagging clearly normal code as a defect, or attributing behavior in a way that contradicts the diff. Only comments where the diff directly proves the description wrong are flagged.
\end{itemize}

This falsification-first design has two consequences. First, it preserves recall: comments that the reflector cannot disprove are retained, even if suspicious, because the SubAgent may have evidence the reflector lacks. Second, it improves precision: comments that are directly contradicted by diff evidence, the most damaging type of hallucination, are removed. The module thus operates as a high-precision, conservative filter rather than an aggressive re-reviewer.

\subsubsection{Filter-Only Design}

The reflection module is strictly a filter: it can remove comments but cannot generate new ones. This design choice ensures that the module's influence is always conservative: reflecting on a set of comments can only reduce false positives, never introduce new false ones. If the LLM's response cannot be parsed, the module fails open: all comments are retained, prioritizing recall over precision in the failure case.

This filter-only design distinguishes \OCR\ from both Reflexion-style systems, where the reflector may rewrite outputs, and from program-analysis validators, which produce their own findings. By restricting the reflector to deletion, the system ensures that the breadth of issues surfaced is determined solely by the SubAgent's exploration, while the reliability of each surfaced comment is independently checked.

\section{Evaluation}

To assess whether \OCR's deterministic engineering philosophy translates into concrete performance gains, we evaluate it on AACR-Bench~\cite{zhang2026aacrbench}, a recently introduced multilingual, repository-level benchmark for automated code review, and compare against two mainstream coding agents that ship built-in review functionality: Claude Code and Codex.

\subsection{Evaluation Setup}

\subsubsection{Benchmark and Metrics}

We evaluate on AACR-Bench~\cite{zhang2026aacrbench}, a multilingual, repository-level benchmark comprising 200 real-world pull requests from 50 open-source repositories across 10 programming languages. Ground truth consists of 1{,}505 review comments validated through an ``\textit{AI-assisted, expert-verified}'' pipeline with three rounds of cross-validation by over 80 senior engineers. Unlike prior benchmarks that provide only diff-level context or rely on noisy raw PR comments, AACR-Bench preserves full repository structure for cross-file exploration and provides expert-verified annotations, offering a reliable basis for evaluating code review systems.

AACR-Bench evaluates generated comments via \emph{semantic matching}: rather than exact string matching, the evaluation pipeline judges whether a generated comment expresses the same concern as a ground-truth comment at the same code location. Semantic equivalence is determined by an LLM judge (Qwen3-235B-A22B-Instruct in this work). Based on the resulting matches, the benchmark computes \textbf{Precision} (fraction of generated comments that semantically match ground truth), \textbf{Recall} (fraction of ground-truth comments matched), and \textbf{SEM-F1} (their harmonic mean), which serves as the primary ranking metric.

\subsubsection{Systems and Models}

We evaluate three code review systems:

\begin{itemize}[leftmargin=*]
  \item \textbf{\OCR} (v1.3.1): our system, configured with its default rule set, curated tool set, and independent reflection module.
  \item \textbf{Claude Code} (v2.1.169, \texttt{/code-review} command): Anthropic's coding agent with built-in review functionality. It employs a general-purpose agent loop without review-specific tools or file-level parallelism.
  \item \textbf{Codex} (v0.140.0, \texttt{/review} command): OpenAI's coding agent with sandboxed repository access, using a single-agent review loop.
\end{itemize}

Each system is evaluated with multiple LLM backends to decouple system design from model capability: \OCR\ is evaluated with all six models (Claude-4.6-Opus, Claude-4.8-Opus, GPT-5.5, GLM-5.1, Qwen3.7-Max, and Deepseek-V4-Pro); Claude Code is evaluated with six models, excluding GPT-5.5; and Codex is evaluated with GPT-5.5 only, due to model compatibility and evaluation cost constraints. All systems share the same LLM configuration, ensuring that observed performance differences reflect system design rather than model advantage.

\begin{table*}[htbp]
  \centering
  \caption{Main results on AACR-Bench. Bold values indicate the best result per model group (row) for each metric. \OCR\ achieves the highest SEM-F1 across all model backends, with consistent advantages in precision, token efficiency, and review time. Avg.~Token and Avg.~Time denote the average token consumption and execution time per sample, respectively. Match/Gen shows matched/generated counts for precision; Match/GT shows matched/ground-truth counts for recall. Ground truth comprises 1{,}505 comments.}
  \label{tab:main-results}
  \begin{tabular}{llrrrrrrr}
    \toprule
    \textbf{Model} & \textbf{System} & \textbf{SEM-F1} & \textbf{Prec.} & \textbf{Match/Gen} & \textbf{Recall} & \textbf{Match/GT} & \textbf{Avg. Token} & \textbf{Avg. Time} \\
    \midrule
    \multirow{2}{*}{Claude-4.6-Opus} & OpenCodeReview & \textbf{25.10\%} & \textbf{33.90\%} & 301/889 & 20.00\% & 301/1505 & \textbf{385K} & \textbf{1m23s} \\
     & Claude Code & 11.57\% & 7.23\% & 435/5980 & \textbf{28.90\%} & 435/1505 & 5{,}664K & 13m06s \\
    \midrule
    \multirow{2}{*}{Qwen3.7-Max} & OpenCodeReview & \textbf{21.20\%} & \textbf{25.20\%} & 276/1096 & 18.30\% & 276/1505 & \textbf{625K} & \textbf{4m41s} \\
     & Claude Code & 12.17\% & 8.23\% & 351/4260 & \textbf{23.37\%} & 351/1505 & 5{,}153K & 8m06s \\
    \midrule
    \multirow{2}{*}{GPT-5.5} & OpenCodeReview & \textbf{21.00\%} & \textbf{32.10\%} & 234/728 & \textbf{15.50\%} & 234/1505 & \textbf{422K} & \textbf{2m51s} \\
     & Codex & 8.36\% & 27.82\% & 74/266 & 4.92\% & 74/1505 & 525K & 2m58s \\
    \midrule
    \multirow{2}{*}{Claude-4.8-Opus} & OpenCodeReview & \textbf{17.90\%} & \textbf{37.80\%} & 176/465 & 11.70\% & 176/1505 & \textbf{352K} & \textbf{1m06s} \\
     & Claude Code & 14.13\% & 15.93\% & 191/1200 & \textbf{12.70\%} & 191/1505 & 2{,}062K & 5m38s \\
    \midrule
    \multirow{2}{*}{Deepseek-V4-Pro} & OpenCodeReview & \textbf{17.90\%} & \textbf{30.60\%} & 191/624 & 12.70\% & 191/1505 & \textbf{394K} & \textbf{6m28s} \\
     & Claude Code & 10.93\% & 8.27\% & 243/2945 & \textbf{16.13\%} & 243/1505 & 5{,}450K & 14m24s \\
    \midrule
    \multirow{2}{*}{GLM-5.1} & OpenCodeReview & \textbf{20.40\%} & \textbf{28.90\%} & 237/820 & 15.70\% & 237/1505 & \textbf{743K} & \textbf{4m11s} \\
     & Claude Code & 11.93\% & 8.37\% & 313/3742 & \textbf{20.80\%} & 313/1505 & 4{,}038K & 14m10s \\
    \bottomrule
  \end{tabular}
\end{table*}

\subsection{Overall Results}

Table~\ref{tab:main-results} presents the full results. \OCR\ achieves the highest SEM-F1 across all six LLM backends, with the top configuration (Claude-4.6-Opus) reaching 25.10\% SEM-F1, 33.90\% precision, and 20.00\% recall. The same model backend under Claude Code achieves only 11.57\% SEM-F1, less than half of \OCR's score, and under Codex, GPT-5.5 achieves merely 8.36\%.
Two patterns emerge from the results:

\paragraph{\textbf{(1) \OCR\ consistently outperforms both baselines across all model backends.}} Across all tested configurations, \OCR's SEM-F1 ranges from 17.90\% to 25.10\%, compared to 10.93\%--14.13\% for Claude Code and 8.36\% for Codex. Notably, the model that ranks first under \OCR\ (Claude-4.6-Opus, 25.10\%) ranks near the bottom under Claude Code (11.57\%), a 2.17$\times$ improvement attributable solely to system design. Even the weakest \OCR\ configuration (Deepseek-V4-Pro, 17.90\%) surpasses the strongest Claude Code configuration (14.13\%), demonstrating that the performance gap is robust to model choice.

\paragraph{\textbf{(2) The performance advantage stems from precision, not recall inflation.}} \OCR's precision ranges from 25.20\% to 37.80\%, versus 7.23\%--15.93\% for Claude Code and 27.82\% for Codex. Claude Code achieves its highest recall (28.90\%) by generating a large volume of comments---4{,}580 across 200 PRs, compared to 465--1{,}096 for \OCR---but at the cost of severe precision degradation (7.23\%). This pattern, where agents compensate for shallow analysis by emitting many low-confidence comments, is precisely the ``\textit{low signal-to-noise ratio}'' problem documented in prior work on deployed code review agents~\cite{chowdhury2026empirical}. \OCR's independent reflection module suppresses this behavior: by filtering comments that are directly contradicted by diff evidence, it maintains high precision without sacrificing issue breadth.

\subsection{Cost Efficiency}

A central claim of this work is that deterministic engineering achieves \emph{better} results at \emph{lower} cost. Table~\ref{tab:main-results} also isolates this dimension by comparing token consumption across systems using the same or comparable model backends.

The results are unambiguous: \textbf{\textit{\OCR\ consumes 5--15$\times$ fewer tokens than Claude Code while achieving 1.3--2.2$\times$ higher SEM-F1}}. The token savings derive directly from three deterministic engineering decisions: (1)~rule-guided dispatch avoids the exploratory overhead of agent-driven file triage; (2)~bounded-output curated tools prevent the ``\textit{token snowball}'' effect~\cite{fan2025sweeffi} that inflates context in free-form agent loops; and (3)~file-level parallelism with per-file context isolation avoids the monolithic context window that centralized agents accumulate. Against Codex, \OCR\ achieves 2.5$\times$ higher SEM-F1 at comparable token cost. Codex operates as a single-agent loop with general-purpose repository access, which may limit its ability to surface deep, cross-file issues.

\begin{figure}[ht]
    \centering
    \includegraphics[width=1.0\linewidth]{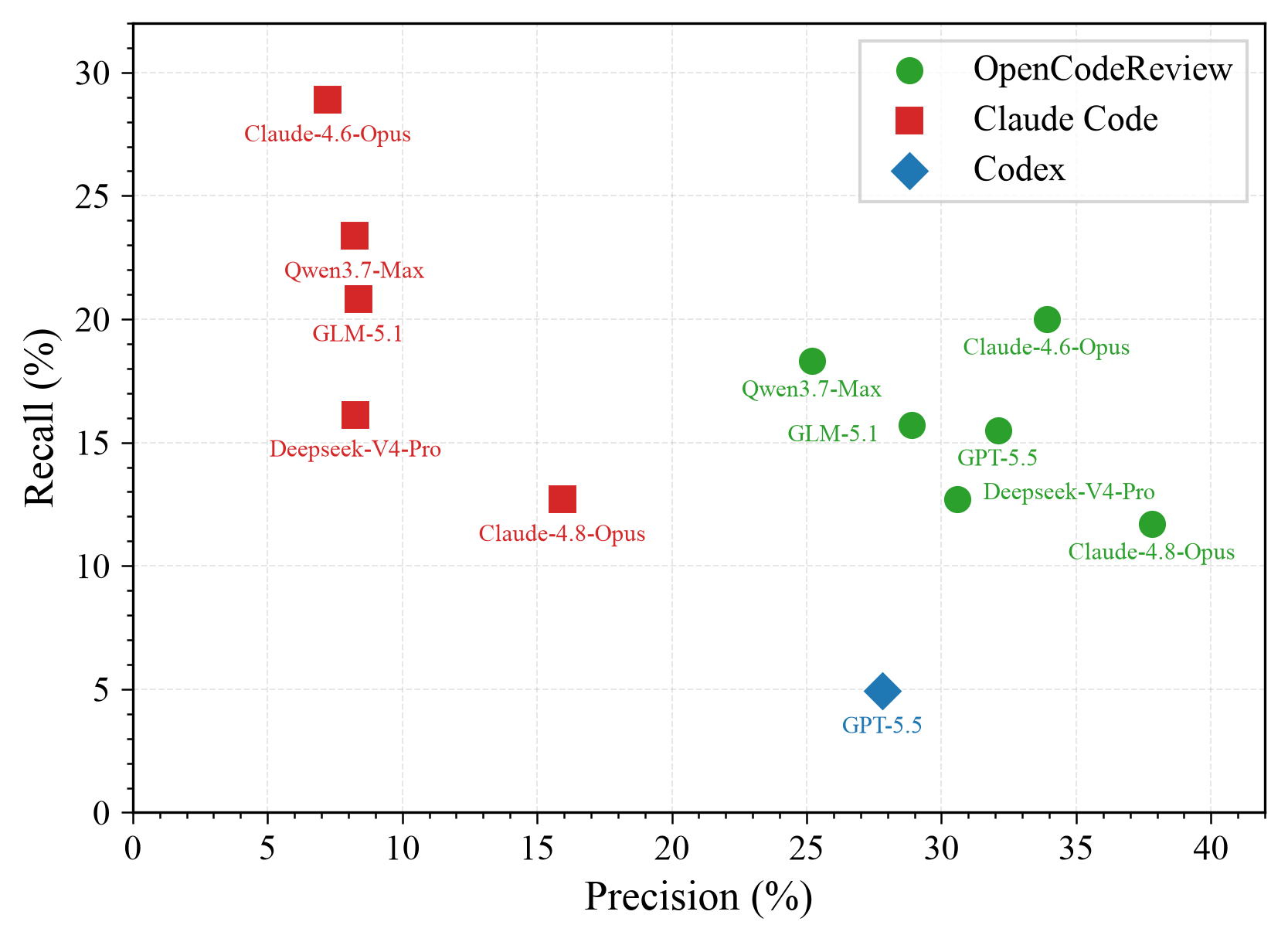}
    \caption{Precision-recall trade-off across all 12 configurations. \OCR\ configurations cluster in the high-precision, moderate-recall region, while Claude Code and Codex occupy distinct regimes with lower precision at comparable or lower recall.}
    \label{fig:precision-recall}
\end{figure}

\subsection{Precision-Recall Analysis}

The precision-recall trade-off is the central tension in automated code review: an aggressive system surfaces many issues (high recall) at the cost of false positives (low precision), while a conservative system generates few but accurate comments (high precision) at the risk of missing genuine issues (low recall). 
Figure~\ref{fig:precision-recall} plots all 12 configurations in this space.

\OCR\ configurations cluster in the high-precision, moderate-recall region (precision 25--38\%, recall 12--20\%). Claude Code configurations split into two regimes: the Claude-4.6-Opus and Qwen3.7-Max variants achieve high recall (23--29\%) but with extremely low precision (7--8\%), while Claude-4.8-Opus achieves moderate precision (16\%) at low recall (13\%). Codex occupies a third regime: very low recall (5\%) with moderate precision (28\%), suggesting that it generates few but relatively accurate comments---likely because its single-agent loop terminates early on most PRs.

\OCR's favorable position in this trade-off is a direct consequence of the independent reflection module. The reflection module's falsification-first design---filtering only comments directly contradicted by diff evidence---preferentially removes false positives without suppressing true positives that reference context beyond the diff. This explains why \OCR\ maintains precision above 25\% even when generating 700--1{,}100 comments across the benchmark, while Claude Code's precision drops to 7--8\% at similar or higher generation volumes.

\section{Discussion}

The evaluation demonstrates that \OCR's deterministic engineering---curated tools, file-level parallelism, and independent reflection---yields consistently better review quality at a fraction of the cost of mainstream coding agents. Beyond the specific numbers, three broader insights emerge:

\paragraph{(1) Determinism as a design principle.} A prevailing assumption in LLM agent research is that more autonomy leads to better outcomes. Our results challenge this: \OCR\ deliberately constrains the agent at three levels---file selection, action space, and output reliability---yet outperforms less constrained agents across all model backends. This resonates with prior findings that tool-interface design materially affects performance~\cite{yang2024sweagent}, unbounded exploration causes the ``\textit{token snowball}'' effect~\cite{fan2025sweeffi}, and vast action spaces introduce non-determinism~\cite{yagubyan2026consistency}. For well-scoped, repetitive tasks like code review, non-determinism is not an inherent property of LLM agents but a design failure that can be engineered away by constraining the action space along task-validated axes.

\paragraph{(2) Information boundaries vs.\ model boundaries for reflection.} The reflection module achieves independence through an asymmetric information boundary rather than a different model. Unlike Reflexion~\cite{shinn2023reflexion} and Self-Refine~\cite{madaan2023selfrefine}, which use the same model with the same context (inheriting self-reinforcing bias), or external validators~\cite{guo2025repoaudit}, which cover only narrow fact types, \OCR's reflector uses the same LLM but with less context (i.e., seeing only the diff) and applies a falsification-first principle. Its effectiveness (precision 25--38\% across six backends) suggests that for reflection, the information boundary may matter more than the model identity, a principle that may generalize to other agent tasks where outputs can be checked against a subset of the evidence.

\paragraph{(3) The cost-quality frontier.} The 5--15$\times$ token savings over Claude Code come from \emph{structuring} exploration, not eliminating it: rule-guided dispatch avoids wasteful file triage, bounded tools cap per-call context growth, and file-level parallelism prevents monolithic context accumulation---all without sacrificing the ability to trace cross-file dependencies. At industrial scale, where review is a primary bottleneck~\cite{sadowski2018modern}, this cost reduction transforms agent-based review from a premium feature to a viable default, suggesting that the cost-quality frontier for LLM agents is not fixed but can be shifted inward through deliberate engineering.

\section{Threats to Validity}

\paragraph{Internal validity.} LLM-based agent systems are inherently non-deterministic~\cite{yagubyan2026consistency}. \OCR\ mitigates this through three design decisions---Rule-Guided Dispatch, Grounded File Review, and Independent Reflection---each injecting determinism at a distinct point in the pipeline. All systems are further evaluated under identical sampling parameters. A related threat concerns the LLM judge used for semantic matching in metrics computation. To mitigate evaluator non-determinism, we run the judge five times per configuration and report the mean metrics as the final results. Since all systems are evaluated by the same matcher under the same protocol, relative comparisons remain valid.

\paragraph{External validity.} While AACR-Bench cannot cover all industrial codebases---particularly domain-specific languages or repositories with unconventional structures---its coverage of 10 programming languages and 50 repositories provides representative diversity for evaluating code review agents. To further strengthen generalizability, we evaluate across six LLM backends spanning four providers. Although only two baselines (Claude Code and Codex) are compared, both represent the current state-of-the-art in industry-shipped coding agents with built-in review functionality, ensuring that the observed results are reliable and meaningful.

\paragraph{Construct validity.} SEM-F1, as a semantic matching metric, may not fully capture the practical value of review comments: factors such as actionability, clarity, and severity are not measured. Conversely, a genuinely useful comment that does not match any ground-truth item would be counted as a false positive. To mitigate this, we adopt the expert-verified ground truth of AACR-Bench, which was validated through three rounds of cross-validation by over 80 senior engineers, substantially reducing the likelihood of missing genuine issues.

\section{Conclusion}

We presented \OCR, an LLM-based code review agent built on the principle of \emph{deterministic engineering for uncertain agents}. \OCR\ injects determinism at three deliberate points: Rule-Guided Dispatch (a multi-layer rule system for file and criterion selection), Grounded File Review (a curated tool set with bounded outputs for constrained exploration), and Independent Reflection (an independent reflection module with an asymmetric information boundary and falsification-first principle).

On AACR-Bench, \OCR\ consistently outperforms mainstream coding agents across six LLM backends, achieving up to 2.17$\times$ higher SEM-F1 while consuming 5--15$\times$ fewer tokens. The evaluation confirms that system design contributes more to review quality than model choice, that structured exploration achieves both better quality and lower cost, and that independence in reflection can be achieved through information boundaries rather than model identity.

Looking forward, we see two promising directions: automating rule discovery from historical review data to reduce manual maintenance, and generalizing the asymmetric information boundary principle to other agent tasks where outputs can be partially verified against a subset of the evidence. We hope that \OCR's open-source release and its deterministic engineering philosophy contribute to making LLM-based code review both reliable and economical for real-world deployment.

\bibliographystyle{ACM-Reference-Format}
\bibliography{main}


\end{document}